\documentclass[aps,prx,twocolumn,superscriptaddress,showpacs,longbibliography]{revtex4-1}
\usepackage{amsmath}
\usepackage{graphicx}
\usepackage{amsfonts}
\usepackage{verbatim}
\usepackage{amssymb}
\usepackage{color}
\usepackage{dsfont}
\usepackage{amsmath} % or simply amstext
\usepackage{hyperref}
\hypersetup{colorlinks = true, urlcolor = blue, linkcolor = blue, citecolor = blue}

\newcommand{\angstrom}{\text{\normalfont\AA}}

\newcommand{\su}{\uparrow} 
\newcommand{\sd}{\downarrow} 
\newcommand{\bpm}{\begin{pmatrix}}
\newcommand{\epm}{\end{pmatrix}}

\newcommand{\nn}{\nonumber \\} 
\newcommand{\tp}{ ^{\intercal} }
\newcommand{\dg}{^{\dagger}}
\newcommand{\hc}{\rm{H.c.}}
\newcommand{\half}{\frac{1}{2}}

\newcommand{\nb}{\bar{n}}

\begin{document}

\title{Josephson current via an isolated Majorana zero mode}
\author{Chun-Xiao Liu}\email{chunxiaoliu62@gmail.com}
\affiliation{Qutech and Kavli Institute of Nanoscience, Delft University of Technology, Delft 2600 GA, The Netherlands}

\author{Bernard van Heck}
\affiliation{Microsoft Quantum Lab Delft, Delft University of Technology, 2600 GA Delft, The Netherlands}

\author{Michael Wimmer}
\affiliation{Qutech and Kavli Institute of Nanoscience, Delft University of Technology, Delft 2600 GA, The Netherlands}

\date{\today}

\begin{abstract}
We study the equilibrium dc Josephson current in a junction between an $s$-wave and a topological superconductor.
Cooper pairs from the $s$-wave superconducting lead can transfer to the topological side either via an unpaired Majorana zero mode localized near the junction, or via the above-gap continuum states.
We find that the Majorana contribution to the supercurrent can be switched on when time-reversal symmetry in the conventional lead is broken, e.g., by an externally applied magnetic field inducing a Zeeman splitting.
Moreover, if the magnetic field has a component in the direction of the effective spin-orbit field, there will be a Majorana-induced anomalous supercurrent at zero phase difference.
These behaviors may serve as a signature characteristic of Majorana zero modes, and is accessible to devices with only superconducting contacts.
\end{abstract}

\maketitle

Majorana zero modes (MZMs) are neutral mid-gap excitations localized at the defects or wire ends of a topological superconductor~\cite{Alicea2012New, Leijnse2012Introduction, Beenakker2013Search, Stanescu2013Majorana, Jiang2013Non, Elliott2015Colloquium, Sato2016Majorana, Sato2017Topological, Aguado2017Majorana, Lutchyn2018Majorana, Zhang2019Next, Frolov2019Quest}.
Due to their robustness against local perturbations and their non-Abelian statistics, MZMs are potential building blocks for topological quantum computation~\cite{Nayak2008Non-Abelian, DasSarma2015Majorana}.
One of the promising candidates for realizing topological superconductivity in solid state physics are heterostructures consisting of a one-dimensional Rashba spin-orbit-coupled semiconductor nanowire and a proximitizing conventional $s$-wave superconductor~\cite{Sau2010Generic, Lutchyn2010Majorana, Oreg2010Helical, Sau2010NonAbelian}.
The application of a large enough Zeeman field parallel to the nanowire can drive the hybrid system into the topological superconducting phase, with MZMs forming at the wire ends. 

\begin{figure}
\begin{center}
\includegraphics[width=\linewidth]{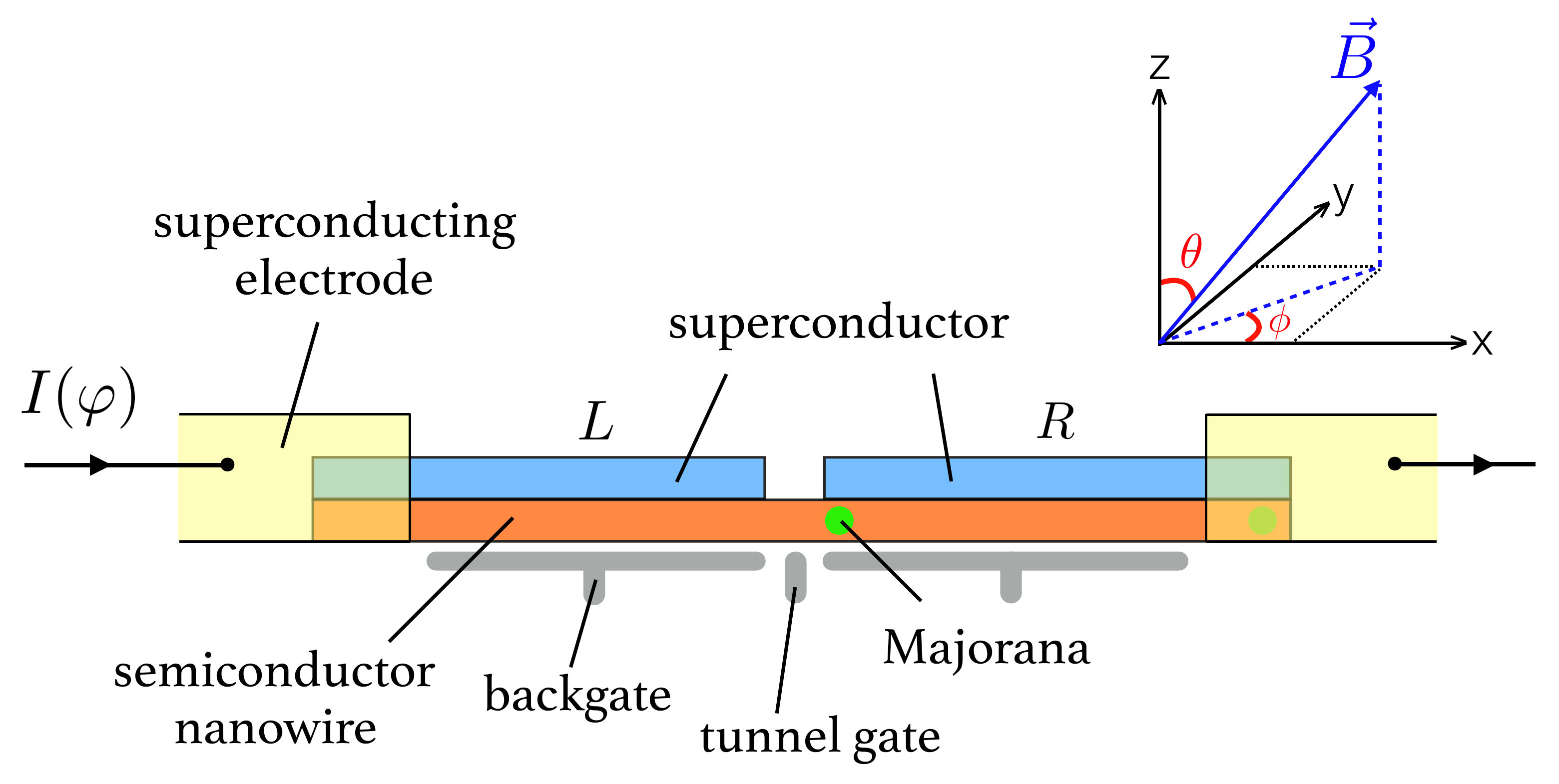}
\end{center}
\caption{Side-view schematic of a Josephson junction between a trivial and a topological superconductor. A semiconducting nanowire (orange) is in proximity with two conventional $s$-wave superconductors (blue) separated by the tunnel junction. An unpaired Majorana zero mode (green dot) can appear near the junction when the right-side hybrid nanowire becomes topological. The chemical potential of the superconductor can be tuned by the backgate (grey line), while the junction transparency can controlled by the tunnel gate (grey dot).
The inset indicates the coordinate axes and direction of magnetic field.}
\label{fig:schematic} 
\end{figure}

So far, most evidence for MZMs comes from tunneling spectroscopy in normal metal-superconductor junctions, in which a MZM gives rise to a zero-bias conductance peak~\cite{Mourik2012Signatures, Das2012Zero, Deng2012Anomalous,Churchill2013Superconductor, Finck2013Anomalous, Albrecht2016Exponential, Chen2017Experimental, Deng2016Majorana, Zhang2017Ballistic, Gul2018Ballistic, Nichele2017Scaling}.
In addition, several proposals have been put forward to probe topological superconductivity with superconducting contacts.
One advantage of a superconducting lead is that quasiparticle poisoning can be mitigated at temperatures smaller than the gap $\Delta_0$, which is beneficial for qubit proposals~\cite{Schrade2018Majorana}.
In a voltage-biased junction between a trivial and a topological superconductor, the MZM will manifest itself as a conductance peak of height $(4-\pi)2e^2/h$ at $eV = \pm \Delta_0$ in the tunneling limit~\cite{Peng2015Robust, Chevallier2016Tomography, Setiawan2017Conductance, Setiawan2017Transport}.
Several works have considered the equilibrium dc Josephson current between a trivial and a topological superconductor (see Fig.~\ref{fig:schematic}) and have established that the Majorana contribution to the supercurrent is negligible~\cite{Zazunov2012Supercurrent,Ioselevich2016Josephson,Zazunov2016Low-energy,Zazunov2018Josephson}.
Corrections arise due to the above-gap quasiparticle contributions, if the nanowire length is short, or if a quantum dot is present between the two leads~\cite{Zazunov2018Josephson, Schuray2018Influence, Cayao2018Finite,  Schrade2018Andreev}.

The existing studies have focused on the case in which time-reversal symmetry (TRS) is present in the trivial superconductor.
In practice, however, a magnetic field has to be applied globally to a device, and thus TRS inside the trivial lead is inevitably broken.
In this work, we explore in detail the consequences of TRS breaking in the trivial lead for the Josephson current.
We show that a finite Zeeman splitting inside the trivial superconductor generates a Majorana-induced supercurrent.
Additionally, if the magnetic field has a component in the direction of the effective spin-orbit field, the MZM induces an anomalous supercurrent, flowing at zero phase difference between the leads.
Thus, under appropriate conditions the dc Josephson current in a trivial-topological superconductor junction can provide observable evidence for MZMs.

\section{Model and method}
The Hamiltonian for the one-dimensional nanowire Josephson junction represented in Fig.~\ref{fig:schematic} is
\begin{align}
H = H_{L} + H_{R} + H_{\rm{tunnel}},
\label{eq:H_tot}
\end{align}
where $H_{L}$ ($H_R$) is the Hamiltonian for the left (right) nanowire lead~\cite{Lutchyn2010Majorana,Oreg2010Helical}:
\begin{align}
\nonumber
H_{j} &= \int dx \,c\dg_{j \sigma}(x)\,[h_j]_{\sigma\sigma'}\,c_{j \sigma'}(x)  + \Delta_0  [c_{j \sd}(x) c_{j \su}(x) +\textrm{h.c.}]\,,\\
h_j&=-\frac{\hbar^2  }{2m^*} \partial^2_x - \mu_{j}  - i \alpha_{j}  \partial_x \sigma_y + \vec{E}_{Z,j} \cdot \vec{\sigma}\,.
\label{eq:H_smsc}
\end{align}
Here $j=L,R$; $c\dg_{j \sigma}(x)$ creates an electron of spin $\sigma$ in lead $j$ at position $x$; $\sigma_{x,y,z}$ are the Pauli matrices acting on the spin space; $m^*$ is the effective mass, $\mu_{j}$ the chemical potential, $\alpha_{j}$ the strength of Rashba spin-orbit coupling with the corresponding spin-orbit field pointing along the $\sigma_y$-direction, $\Delta_0$ the proximity-induced superconducting gap and $\vec{E}_{Z,j} = \half g \mu_B \vec{B}_j$ the Zeeman field due to the applied magnetic field.
We have defined parameters separately for the left and right lead, which will allow us to consider different physical scenarios in what follows.
We will always assume that the chemical potential in the left lead is set to a high value $\mu_L\gg \Delta_0$, such that the left lead is in the topologically trivial regime.
For numerical results, the continuum Hamiltonian in Eq.~\eqref{eq:H_smsc} needs to be discretized into a tight-binding model~\cite{Lin2012Zero, Liu2017Role}.
When doing so, we always take the size $L$ of left and right leads to be large enough that finite size effects (e.g., Majorana overlap) play no essential role.

The tunnel Hamiltonian is given by
\begin{align}
H_{\rm{tunnel}} = -t e^{i\varphi/2} \sum_{\sigma = \su \sd} c\dg_{R\sigma}(x_R) c_{L\sigma}(x_L) + \textrm{h.c.},
\label{eq:H_tunn}
\end{align}
and describes spin-conserving single electron tunneling occurring at a point contact connecting the left lead (ending at $x=x_L$) to the right lead (beginning at $x=x_R$).
Here, $\varphi$ is the phase difference between the leads, and $t$ is the tunneling strength, which is associated with the normal conductance by $G_N=4 \pi e^2 t^2 \nu_L \nu_R / \hbar$, with $\nu_{L,R}$ being the normal density of states at the Fermi surface of left and right lead.

In the tunneling limit $t \ll \Delta_0$, which can always be reached by tuning the tunnel gate, second-order perturbation theory yields the zero-temperature current-phase relation of the junction~\cite{Ambegaokar1963Tunneling,supp},
\begin{align}
I(\varphi) = I_c \sin ( \varphi + \varphi_0)\,.
\label{eq:cpr}
\end{align}
The critical current $I_c = 4et^2  |\mathcal{A}| / \hbar$ and the phase shift $\varphi_0=\arg (\mathcal{A})$ are determined by the amplitude $\mathcal{A}$ of Cooper pair transfer from left to right.
The latter is a sum over all possible intermediate states with a quasiparticle in each lead,
\begin{align}
\mathcal{A} =  \sum_{\substack{nm\\\eta\sigma=\su\sd}}\frac{u^*_{Ln \eta}(x_L)\,v_{Ln \sigma}(x_L)\,u_{Rm \sigma}(x_R)\,v^*_{Rm \eta}(x_R)}{E_{Ln} + E_{Rm}}.
\label{eq:A}
\end{align}
Here $E_{j n} \geq 0$ is the energy of the $n$-th Bogoliubov quasiparticle excitation in lead $j$, with Nambu wave function $[u_{j n \su}(x), u_{j n \sd}(x), v_{j n \su}(x), v_{j n \sd}(x)]\tp$.
When the right lead is in the topological phase, we can separate the amplitude into two parts $\mathcal{A} = \mathcal{A}^{M} + \mathcal{A}^{\text{cont}}$, depending on whether the intermediate state involves an isolated MZM ($E_{Rm}=0$) or an excited quasiparticle state in the continuum ($E_{Rm}>0$).
At zero field, Eq.~\eqref{eq:A} yields the classical Ambegaokar-Baratoff relation $I_{c0} = (\pi/2e) G_N \Delta_0$~\cite{Ambegaokar1963Tunneling}.
When $\varphi_0 \neq 0$ or $\pi$, an anomalous supercurrent  $I_\textrm{an} =I_c \sin ( \varphi_0)$ flows at $\varphi=0$.

\section{Majorana-induced supercurrent}
We now focus on a physical scenario that illustrates the joint role of the MZM and TRS breaking in the left lead in generating a supercurrent.
Namely, we consider the case in which the parameters of $H_R$ are fixed in the topological regime, i.e., the strength of spin-orbit coupling is finite $\alpha_R > 0$, and the Zeeman field is larger than the critical value, $|\vec{E}_{Z,R}| > \sqrt{\Delta^2_0 + \mu^2_R}$.
Under these conditions, and provided the wire is long enough, there will be an unpaired MZM with particle-hole symmetric wave function $[\xi_\su(x), \xi_\sd(x), \xi^*_\su(x), \xi^*_\sd(x)]\tp$ exponentially localized at $x=x_R$.
At the same time, we assume that the left lead is subject to a Zeeman field pointing in an arbitrary direction, possibly different from that of $\vec{E}_{Z,R}$, parametrized by angles $\theta$ and $\phi$ (Fig.~\ref{fig:schematic}) so that $\vec{E}_{Z, L} = E_{Z,L}\,(\sin\theta\cos\phi, \sin\theta\sin\phi, \cos\theta)$.
We further assume that the left lead has no spin-orbit coupling, $\alpha_L=0$, and that $E_{Z,L}<\Delta_0$ to guarantee a finite energy gap.
Under these conditions, the amplitude of Cooper pair transfer via the MZM is~\cite{supp}
\begin{align}
\mathcal{A}^M =  \nu_L   f( \frac{E_{Z,L}}{\Delta_0} ) \left[ ( \xi^2_{\sd} e^{i\phi} -  \xi^2_{\su} e^{-i\phi} ) \sin \theta + 2\xi_{\su} \xi_{\sd} \cos \theta \right]
\label{eq:amplitude_M}
\end{align}
with $ f(x) = \frac{\arcsin(x)}{2\sqrt{1-x^2}}$ \footnote{Note that the divergence of $f(x)$ for $x\to 1$ is not physical and indicates the breakdown of perturbation theory as the gap closes in the left lead.}.
Equation~\eqref{eq:amplitude_M} is the central result of our work and deserves several comments.

First, if $E_{Z,L}=0$, $\mathcal{A}^M=0$ and the Majorana-induced supercurrent is blockaded~\cite{Zazunov2012Supercurrent}.
Although Eq.~\eqref{eq:amplitude_M} assumes no spin-orbit coupling in the trivial SC lead, the blockade of the Majorana-induced supercurrent is more general and it relies on the presence of TRS in the left lead.
In particular, it holds in the presence of spin-orbit coupling as well as non-magnetic disorder, as we derive in an Appendix~\cite{supp}.
A finite $E_{Z,L}$ however breaks TRS in the left lead, and according to Eq.~\eqref{eq:amplitude_M} a supercurrent can flow via the MZM.
The magnitude of $\mathcal{A}^M$ increases linearly for a small Zeeman field, $\mathcal{A}^M\propto E_{Z,L}/\Delta_0$ for $E_{Z,L}\ll\Delta_0$.

Second, the magnitude of the supercurrent also depends crucially on the direction of $\vec{E}_{Z,L}$, a fact which can be understood as follows.
On one hand, because Cooper pairs in the left lead have zero angular momentum, they are composed by two electrons with opposite spin polarizations along the direction dictated by $\vec{E}_{Z,L}$.
The two paired electrons must both tunnel through the MZM in order for $\mathcal{A}^M$ to be finite.
However, the MZM has its own spin polarization - i.e. the orientation along the Bloch sphere associated with the spinor $[\xi_{\su}(x_R), \xi_{\sd}(x_R)]\tp$ - and therefore acts as a spin filter.
Thus, if $\vec{E}_{Z,L}$ is parallel (or anti-parallel) to the spin polarization of the Majorana wave function, the supercurrent will vanish.

Third, the amplitude $\mathcal{A}^M$ is in general complex, which means that the MZM can contribute to an anomalous supercurrent.
Note that the Majorana wave function components $\xi_\sigma$ are real if the Zeeman field in the right lead has no component along the spin-orbit field direction $y$~\cite{Tewari2012Topological}.
In this case, the phase $\varphi^M_0=\arg(\mathcal{A}^M)$ of the amplitude is controlled only by the direction of $\vec{E}_{Z,L}$.

We can illustrate the previous points with simple limits of Eq.~\eqref{eq:amplitude_M}.
Consider for instance the case in which $\vec{E}_{Z,L}$ lies in the $xz$ plane, i.e., $\vec{E}_{Z,L}\cdot\vec{\sigma} = E_Z(\cos\theta\,\sigma_z+\sin\theta\sigma_x)$ [$\phi=0$ in Eq.~\eqref{eq:amplitude_M}] while $\vec{E}_{Z,R}$ points along the wire. Then,
\begin{align}
&\mathcal{A}^M = \nu_L   f( \frac{E_{Z,L}}{\Delta_0} ) \left[  ( \xi^2_{\sd} -  \xi^2_{\su} ) \sin \theta + 2 \xi_{\su} \xi_{\sd} \cos \theta \right],
\label{eq:xz_plane}
\end{align}
with real wave functions $\xi_\sigma$.
We see that the $\mathcal{A}^M$ vanishes if $\theta=\pi/2$ and the MZM is polarized along the $x$-axis ($\xi^2_{\su}=\xi_{\sd}^2$), and likewise if $\theta=0$ and the MZM is spin-polarized along the $z$-axis ($\xi_{\su}\xi_{\sd}=0$).
Furthermore $\mathcal{A}^M$ is real and thus the MZM does not induce any anomalous supercurrent.
The fundamental reason for the absence of phase shift ($\varphi_0=0$) is that the one-dimensional semiconductor-superconductor nanowire has an additional chiral symmetry (reality of the BdG Hamiltonian) when the applied Zeeman field is perpendicular to the Rashba spin-orbit field~\cite{Tewari2012Topological, Rasmussen2016Effects}.
By contrast, once the Zeeman field has some component along the spin-orbit field ($\sigma_y$), the chiral symmetry is broken and the phase shift becomes finite, as indicated by Eq.~\eqref{eq:amplitude_M} with $\phi \neq 0$. In particular, when the Zeeman field inside the trivial lead is parallel to the $y$-axis ($\theta=\pi/2, \phi=\pi/2$), i.e., $\vec{E}_{Z,L} \cdot \vec{\sigma} = E_{Z, L}\sigma_y$, we have
\begin{align}
\mathcal{A}^M = i \nu_L   f( \frac{E_{Z,L}}{\Delta_0} ) ( \xi^2_{\sd} +  \xi^2_{\su}) \quad\Rightarrow \quad \varphi^M_0 = \pi/2.
\label{eq:y_axis}
\end{align}

Equations~\eqref{eq:amplitude_M}, \eqref{eq:xz_plane}, and \eqref{eq:y_axis}  show that a Zeeman field inside the trivial lead can generate Majorana-induced supercurrent in a trivial-topological superconductor junction, and furthermore that it can lead to anomalous supercurrent.
Although we have assumed zero spin-orbit coupling inside the trivial lead to derive a closed form of Eq.~\eqref{eq:amplitude_M}, such an assumption is not essential, and all the qualitative behavior of $\mathcal{A}^M$ will carry over for finite $\alpha_L$, as we will show in the following. 
Note that even though the Majorana-induced supercurrent may be zero, in general the junction will have a finite supercurrent due to the contribution from the above-gap continuum states in the topological superconductor.
We now resort to numerical simulations in order to compute the total critical current; we will also use this opportunity to relax the simplifying assumptions of the analytical calculation.

\begin{figure}
\begin{center}
\includegraphics[width=\linewidth]{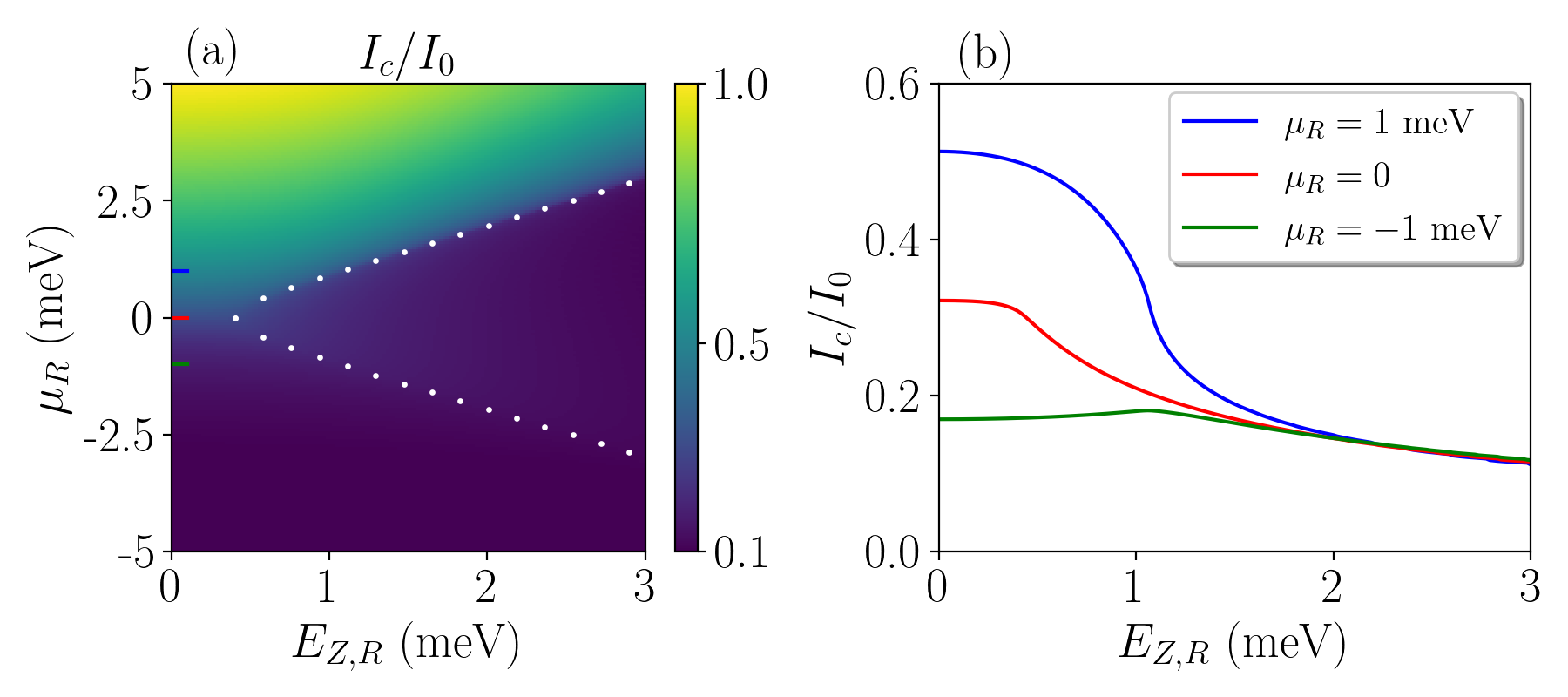}
\end{center}
\caption{Critical Josephson current in a junction between a time-reversal invariant trivial superconductor and a Majorana nanowire lead, with an external Zeeman field ($E_{Z,R} \sigma_x$) applied only inside the Majorana nanowire. (a) $I_c$ as a function of $E_{Z,R}$ and $\mu_R$, with white dots representing the phase boundary ($E^2_{Z,R} = \mu^2_R + \Delta^2_0$ ) of the Majorana nanowire. (b) Linecuts of $I_c$ at fixed values of $\mu_R$. Here the supercurrent is due to above-gap continuum states, without any Majorana contribution. When Zeeman field is larger than the critical value ($E_{Z,R} \gtrsim \sqrt{ \mu^2_R + \Delta^2_0}$), $I_c$ plunges with the field strength, indicating the topological quantum phase transition of the Majorana nanowire.}
\label{fig:Ic_TRI} 
\end{figure}

\section{Numerical simulations}
To get $I(\varphi)$ numerically, we first calculate the eigenenergies and eigenfunctions for the discretized models of the leads in Eq.~\eqref{eq:H_smsc} using the kwant package~\cite{kwant}, and then plug them into Eq.~\eqref{eq:cpr} and ~\eqref{eq:A}.
The parameters are chosen as $m^*= 0.015m_e$, $\alpha_L = \alpha_R=0.5$eV$\angstrom$ ($E_{\text{so}} = \half m^*\alpha^2/\hbar^2 \approx 0.25$meV), $\Delta_0 = 0.4$meV, $\mu_L = 5$meV, and $L=3.5\mu$m.
In the figures, we adopt the value of critical current at $E_{Z,L} = E_{Z,R}=0$, and $\mu_R=5$meV as a unit of supercurrent $I_0$.

Figure~\ref{fig:Ic_TRI} shows the supercurrent in a junction between a time-reversal invariant trivial superconducting lead and a Majorana nanowire lead.
A Zeeman field along the wire axis is applied only inside the Majorana nanowire lead (i.e., $E_{Z,R}\sigma_x$ and $E_{Z,L}=0$).
All the supercurrent originates from above-gap contribution to Eq.~\eqref{eq:A}, while the Majorana-induced supercurrent is blockaded due to the TRS in the trivial lead.
Figure~\ref{fig:Ic_TRI}(a) shows the critical current $I_c$ as a function of Zeeman field $E_{Z,R}$ and the chemical potential $\mu_R$ of the right lead.
In general, the supercurrent is larger when the Majorana nanowire has positive $\mu_R$ and is in the topologically trivial phase $E_{Z,R} < \sqrt{\mu^2_R + \Delta^2_0}$.
Figure~\ref{fig:Ic_TRI}(b) shows linecuts of critical current as a function of Zeeman field at fixed values of chemical potential $\mu_R=0, \pm 1$meV.
The critical current decreases monotonically with the field strength (except for negative $\mu_R$ where the electron density is increased by incressing the Zeeman splitting), and in particular, $I_c$ plunges abruptly near the critical Zeeman field, indicating the topological quantum phase transition of the Majorana nanowire.
These results reproduce previous findings of Ref.~\onlinecite{Zazunov2018Josephson}.

\begin{figure}[tbp]
\begin{center}
\includegraphics[width=\linewidth]{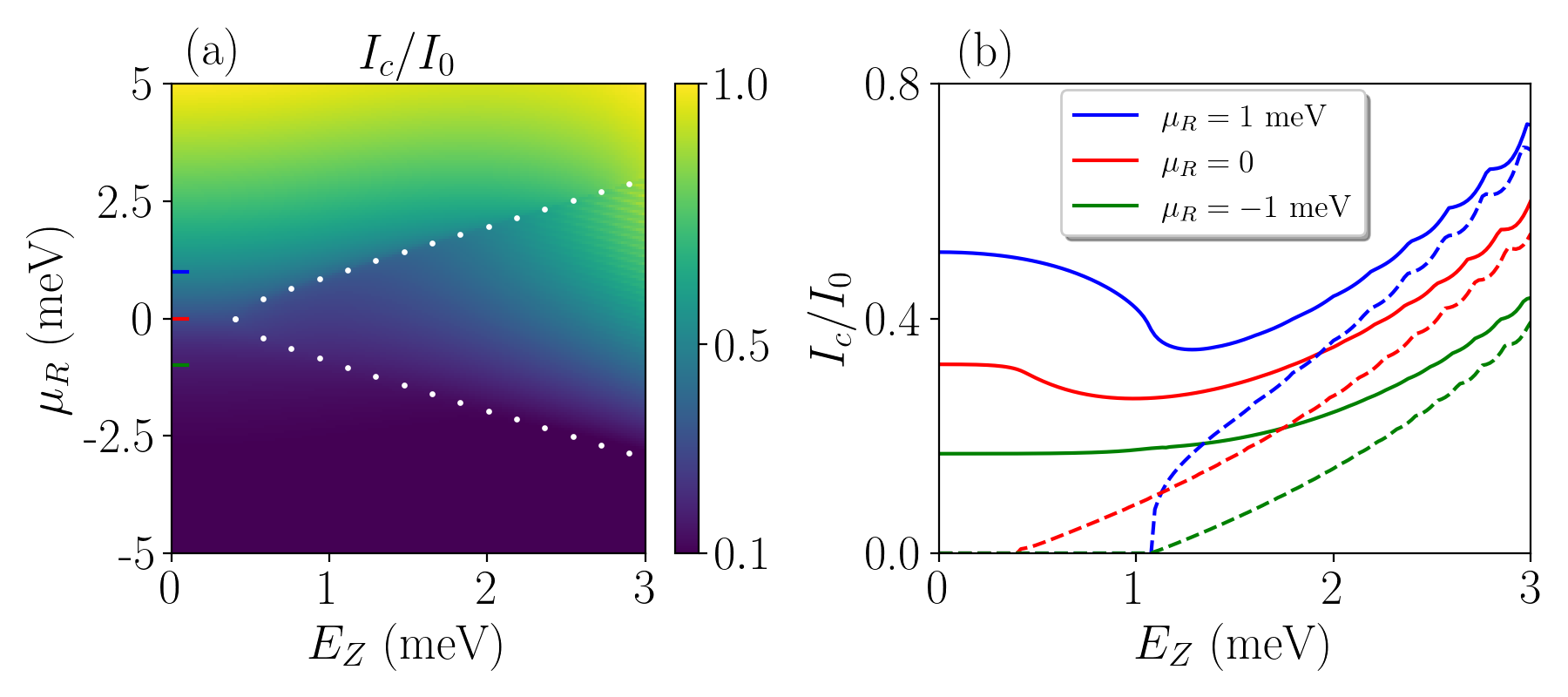}
\end{center}
\caption{Critical current in a junction between a trivial superconductor and a Majorana nanowire lead, with an external Zeeman field applied equally in both leads ($E_Z\sigma_x$). (a) $I_c$ as a function of $E_{Z}$ and $\mu_R$, with white dots representing the phase boundary ($E^2_{Z} = \mu^2_R + \Delta^2_0$) of the Majorana nanowire. (b) Linecuts of $I_c$ at fixed values of $\mu_R$. Solid lines are the total critical current from both MZM and continuum states, while dashed lines are critical current due to MZM only.}
\label{fig:Ic_TRSB} 
\end{figure}

In Fig.~\ref{fig:Ic_TRSB}, we show the calculated supercurrent in a junction between a trivial superconductor and a Majorana nanowire lead, with a Zeeman field along the wire axis being applied globally ($E_Z\sigma_x$ in both leads).
In contrast with Fig.~\ref{fig:Ic_TRI}, now the supercurrent in the topological regime ($E_Z > E_{Zc}$) is also large as shown in Fig.~\ref{fig:Ic_TRSB}(a), because the Majorana-induced supercurrent is finite when TRS in the trivial lead is broken.
Figure~\ref{fig:Ic_TRSB}(b) shows several linecuts of $I_c$ as function of $E_Z$ at fixed values of $\mu_R$ (solid lines).
Instead of monotonically decreasing, the critical current now increases with the field when the Majorana nanowire enters the topologically nontrivial phase.
As shown by the dashed lines in Fig.~\ref{fig:Ic_TRSB}(b), the dominant contribution to $I_c$ deep into the topological phase comes from the MZM, which is consistent with Eq.~\eqref{eq:xz_plane}.
The oscillations of $I_c$ at large $E_Z$ are due to the onset of a finite overlap between two MZMs at the opposite ends of the nanowire.

\begin{figure}
\begin{center}
\includegraphics[width=\linewidth]{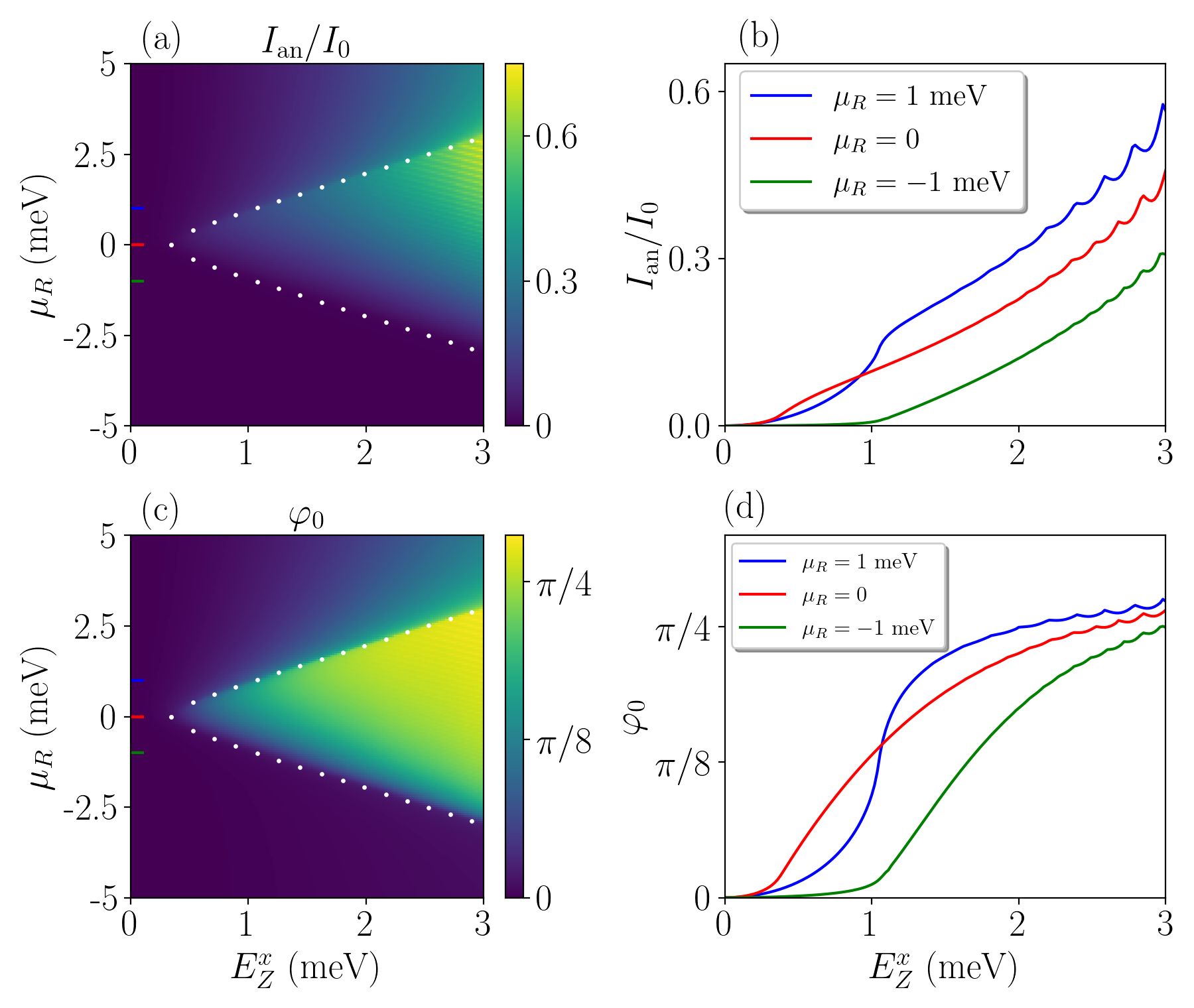}
\end{center}
\caption{Anomalous supercurrent ($I_\textrm{an}$) and phase shift ($\varphi_0$) for the Josephson junction with a Zeeman field applied globally in the form of $\vec{E}_Z \cdot \vec{\sigma} = E^x_Z \sigma_x + E^y_Z \sigma_y$. The component along the spin-orbit field is fixed at $E^y_Z=0.2$meV. (a) and (c) $I_\textrm{an}$ and $\varphi_0$ as a function of $\mu_R$ and $E^x_Z$. Their values in the topological regime are much larger than in the trivial regime due to the Majorana contribution. Here white dots represent the phase boundary $(E^x_Z)^2 + (E^y_Z)^2 = \mu^2_R + \Delta^2_0$. (b) and (d) Linecuts of $I_\textrm{an}$ and $\varphi_0$ for fixed values of $\mu_R$. Note that $I_\textrm{an}$ and $\varphi_0$ increases abruptly near the topological transition.}
\label{fig:I_an} 
\end{figure}

Finally, we consider a Josephson junction for which the Zeeman field is applied globally and has a nonzero component along the spin-orbit field~\cite{Bommer2019SpinOrbit, Liu2019Conductance}.
Namely, the Zeeman field takes the form of $\vec{E}_Z \cdot \vec{\sigma} = E^x_Z \sigma_x + E^y_Z \sigma_y$ in both leads, with the $\sigma_y$-component being fixed at $E^y_Z=0.2$meV $< \Delta_0$.
Here, $E^y_Z \sigma_y$ term breaks the chiral symmetry of the junction leads, and thereby can induce anomalous supercurrent.
Figure~\ref{fig:I_an} shows the corresponding anomalous supercurrent $I_\textrm{an} = I_c \sin(\varphi_0)$ and phase shift $\varphi_0$ in the junction.
As shown in Figs.~\ref{fig:I_an}(a) and~\ref{fig:I_an}(c), $I_\textrm{an}$ and $\varphi_0$ are prominently large inside the topologically nontrivial regime [$(E^x_Z)^2 + (E^y_Z)^2 > \mu^2_R + \Delta^2_0$] due to the Majorana contribution.
Figures~\ref{fig:I_an}(b) and~\ref{fig:I_an}(d) show linecuts of $I_\textrm{an}$ and $\varphi_0$ at fixed $\mu_R$.
For zero and negative $\mu_R$ (red and green curves), $I_\textrm{an}$ and $\varphi_0$ becomes finite only when the Majorana nanowire enters the topological phase, because the continuum states induced supercurrent is negligible.
In contrast, for positive $\mu_R$ (blue curves), the amplitude of $I_\textrm{an}$ and $\varphi_0$ do not vanish in the topologically trivial regime owing to finite contributions from the continuum states. However, a kink in $I_\textrm{an}$ or an abrupt increase of $\varphi_0$ shows up near the critical Zeeman field, signaling the topological quantum phase transition.

\section{Discussion}
We have studied the Josephson current in a nanowire junction between a trivial and a topological superconductor.
We find that a finite Zeeman field in the trivial lead can switch on the Majorana-induced supercurrent and enhance the critical supercurrent.
Furthermore, if the Zeeman field has a component along the spin-orbit field, a MZM can be signaled by the anomalous supercurrent or phase shift.
Thereby a measurement of the dc Josephson current in a trivial-topological superconductor junction as a function of magnetic field and chemical potential could provide compelling evidence for MZMs.
In this respect, our findings parallel those previously obtained for junctions of two topological superconductors~\cite{Jiang2011Unconventional, Badiane2011Nonequilibrium, San-Jose2012ac, Pikulin2012Phenomenology, SanJose2014Mapping, Nesterov2016Anomalous, Marra2016Signatures, Cayao2017Majorana,Murthy2019}.
However, the current proposal simplifies considerably the tuning process of the device by requiring only one superconductor lead to be in the topological phase.
In particular, our proposal provides a way for tuning up Majorana superconducting qubit devices~\cite{Schrade2018Majorana} without the need to add additional probes to their proposed design.

Finally, a few limitations in our work need to be mentioned.
For example, although the Majorana signatures proposed in this work are robust against weak non-magnetic disorders, it may be hard to distinguish Majoranas from smooth-potential-induced low-energy Andreev bound states or quasi-Majoranas within this proposal~\cite{Kells2012Near, Prada2012Transport, Liu2017Andreev, Liu2018Distinguishing, Moore2018Majorana, Reeg2018Zero, Vuik2019Reproducing}, because the supercurrent is induced only by local tunneling processes.
Also, the orbital effect of magnetic field and multi-subband effects are not discussed, which requires model study for two- or three-dimensional systems~\cite{Lutchyn2011Search,Nijholt2016Orbital,Dmytruk2018Suppression, Winkler2019Unified}.
Finally, the point contact model for the tunneling junction may be too simple to describe the coupling between the two segments of the wire.
We thus expect that this work would motivate more investigations on similar Josephson junction devices at a more realistic level.

The data set and code for generating the figures in the work can be found in Ref.~\cite{Liu2020Josephson_zenodo}.

\begin{acknowledgements}
We would like to thank Ji-Yin Wang and F. Setiawan for stimulating discussions, and A. L. R. Manesco for useful comments on the manuscript. This work was supported by a subsidy for top consortia for knowledge and innovation (TKl toeslag) by the Dutch ministry of economic affairs, by the Netherlands Or- ganisation for Scientific Research (NWO/OCW) through VIDI grant 680-47-537, and by support from Microsoft Research.\\
\end{acknowledgements}

\textit{Author contributions}--This project was initiated by C.-X.L. C.-X.L. and B.vH. performed the analytical calculations, and the numerical simulations were performed by C.-X.L. M.W. and B.vH. supervised the project. All authors discussed the results and contributed to writing the manuscript.

\bibliography{references}

\appendix

\onecolumngrid
%\vspace{1cm}
%\begin{center}
%{\bf\large Appendix}
%\end{center}
%\vspace{0.5cm}

\setcounter{secnumdepth}{3}
\setcounter{equation}{0}
\setcounter{figure}{0}
\renewcommand{\theequation}{S-\arabic{equation}}
\renewcommand{\thefigure}{S\arabic{figure}}
\renewcommand\figurename{Supplementary Figure}
\renewcommand\tablename{Supplementary Table}
\newcommand\Scite[1]{[S\citealp{#1}]}
\newcommand\Scit[1]{S\citealp{#1}}

\makeatletter \renewcommand\@biblabel[1]{[S#1]} \makeatother
%%%%%%%%%%%%%%%%%%%%%%%%%%%%%%%%%%
% The supplementary text starts here
%%%%%%%%%%%%%%%%%%%%%%%%%%%%%%%%%%

\section{Derivation of the general formula for supercurrent}\label{app_sec:supercurrent}
We derive the formula for the supercurrent through the Josephson junction in the tunneling limit, as shown in Eqs.~\eqref{eq:cpr} and ~\eqref{eq:A}. The electron operators in the tunneling Hamiltonian $H_{\text{tunn}}$ can be expanded in terms of Bogoliubov quasiparticle operators in the corresponding superconducting lead as
\begin{align}
c_{j \sigma}(x) = \sum_{n} u_{jn\sigma}(x) \Gamma_{jn} + v^*_{j n \sigma}(x) \Gamma\dg_{jn},
\end{align}
where $j=L/R$; $\Gamma\dg_{jn}$ creates a Bogoliubov quasiparticle with excitation energy $E_{jn}$ in lead $j$, and $[u_{j n \su}(x), u_{j n \sd}(x),v_{j n \su}(x), v_{j n \sd}(x) ]\tp$  is the corresponding Nambu wave function. Using the perturbation theory, the phase-dependent part of the ground-state energy is
\begin{align}
E_{gs}(\varphi) &= - \langle \Omega_0 | H_{\rm{tunnel}} H^{-1}_0 H_{\rm{tunnel}}  | \Omega_0 \rangle \nn
&= -t^2e^{i\varphi} \sum_{\sigma, \eta= \su \sd} \langle \Omega_0 | \Big[ c\dg_{R\sigma}(x_R) c_{L\sigma}(x_L) \Big] H^{-1}_0 \Big[ c\dg_{R\sigma}(x_R) c_{L\sigma}(x_L) \Big]  | \Omega_0 \rangle + \hc \nn
&= -t^2e^{i\varphi} \sum_{\sigma, \eta= \su \sd} \sum_{n, m} \langle \Omega_0 | \Big[ v_{Rm \eta}(x_R) \Gamma_{Rm} u_{Ln\eta}(x_L) \Gamma_{Ln} \Big] H^{-1}_0 \Big[ u^*_{Rm\sigma}(x_R) \Gamma\dg_{Rm} v^*_{L n \sigma}(x_L) \Gamma\dg_{Ln} \Big]  | \Omega_0 \rangle + \hc \nn
&= t^2e^{i\varphi} \sum_{\sigma, \eta= \su \sd} \sum_{n, m} \frac{  u_{Ln\eta}(x_L) v^*_{L n \sigma}(x_L) v_{Rm \eta}(x_R) u^*_{Rm\sigma}(x_R)}{E_{Ln} + E_{Rm}} + \hc.
\end{align}
If we further define the Cooper pair transfer amplitude $\mathcal{A}$ as
\begin{align}
\mathcal{A} =   \sum_{\eta, \sigma = \su \sd} \sum_{n,m} \frac{  u_{Ln\eta}(x_L) v^*_{L n \sigma}(x_L) v_{Rm \eta}(x_R) u^*_{Rm\sigma}(x_R) }{E_{Ln} + E_{Rm}},
\end{align}
the ground-state energy becomes
\begin{align}
E_{gs}(\varphi) &=  t^2   \Big( e^{i\varphi} \mathcal{A} + e^{-i\varphi} \mathcal{A}^* \Big) \nn
&= 2 t^2 | \mathcal{A} | \cos( \varphi + \varphi_0 ),
\end{align}
where $\varphi_0 = \arg(\mathcal{A})$. Thereby the current-phase relation is
\begin{align}
I(\varphi) &= -\frac{2e}{\hbar} \frac{\partial E_{gs}(\varphi)}{\partial \varphi} = \frac{4et^2}{\hbar} | \mathcal{A} | \sin( \varphi + \varphi_0 ) = I_c \sin( \varphi + \varphi_0 ),
\end{align}
where $I_c = 4et^2 | \mathcal{A} | / \hbar$.

\section{Transfer amplitude for finite Zeeman field inside the trivial lead}\label{app_sec:Zeeman}
The Hamiltonian for the Josephson junction we consider is 
\begin{align}
H_{L} &= \sum_{\sigma = \su \sd} \int dx c\dg_{L \sigma} \Big ( \frac{ -\partial^2_x }{2m^*} - \mu_{L} + \vec{E}_{Z,L} \cdot \vec{\sigma}  \Big)_{\sigma \sigma'} c_{L \sigma'} + \Delta_0  \int dx ( c_{L \sd} c_{L \su} + c\dg_{L \su} c\dg_{L \sd}  ) \nn
&= \sum_{\sigma, \sigma' = +/-} \int dx c\dg_{L \sigma} \Big ( \frac{ -\partial^2_x }{2m^*} - \mu_{L} + E_{Z,L} \tilde{\sigma}_z  \Big)_{\sigma \sigma'} c_{L \sigma'}  + \Delta_0\int dx  ( c_{L -} c_{L +} + c\dg_{L +} c\dg_{L -}  ), \nn
H_{\rm{tunn}}&= -t e^{i\varphi/2} \sum_{\sigma = +/-} c\dg_{R\sigma}(x_R) c_{L\sigma}(x_L) + \hc.
\end{align}
Here we rotate the spin basis from $| \su \rangle, | \sd \rangle$ to $| + \rangle, | - \rangle$, where $| + \rangle, | - \rangle$ are the eigenstates of $\vec{E}_{Z,L} \cdot \vec{\sigma} $. Thus the Zeeman term becomes diagonal in the rotated basis, i.e., $\vec{E}_{Z,L} \cdot \vec{\sigma} \to E_{Z,L} \tilde{\sigma}_z$. On the other hand, the electron operators can be expanded as
\begin{align}
& c_{L+}(x_L)  = \sum_k \Big( \tilde{u}_n \Gamma_{n, +} - \tilde{v}_n \Gamma\dg_{\bar{n}, -} \Big), \nn
& c_{L-}(x_L) = \sum_k \Big( \tilde{u}_n \Gamma_{\bar{n}, -} + \tilde{v}_n \Gamma\dg_{n, +} \Big), \nn
& c_{R\sigma}(x_R) = \xi_{\sigma}(x_R) \gamma,
\end{align}
where $\Gamma\dg_{n, +}$ and $\Gamma\dg_{\bar{n}, -}$ create the Bogoliubov quasiparticles of excitation energy $E_{n \pm} =  \sqrt{\varepsilon^2_n + \Delta^2_0} \pm E_{Z,L}$ with $\varepsilon_n = \xi_n - \mu_L$. $\tilde{u}_n, \tilde{v}_n$ are BCS coherence factors with $\tilde{u}^2_n = \half + \frac{\varepsilon_n}{2\sqrt{\varepsilon^2_n + \Delta^2_0}} = 1 - \tilde{v}^2_n$. Substituting them into Eq.~\eqref{eq:cpr}, we get
\begin{align}
\mathcal{A}^M &= \xi_{+}(x_R) \xi_{-} (x_R) \sum_n \tilde{u}_n \tilde{v}_n \left( \frac{1}{\sqrt{\varepsilon^2_n + \Delta^2_0} - E_{Z,L}} - \frac{1}{\sqrt{\varepsilon^2_n + \Delta^2_0} + E_{Z,L}} \right) \nn
& = \xi_{+}(x_R) \xi_{-} (x_R) \nu_L \int d\varepsilon_n \frac{ \Delta_0 E_{Z,L}}{ \sqrt{\varepsilon^2_n + \Delta^2_0} ( \varepsilon^2_n + \Delta^2_0 - E^2_{Z,L} ) }  \nn
&=  \nu_L  \xi_{+}(x_R) \xi_{-} (x_R) \frac{\arcsin( E_{Z,L}/\Delta_0)}{ \sqrt{\Delta^2_0 - E^2_{Z,L}} },
\end{align}
for $E_{Z,L} < \Delta_0$. Finally we rotate the spin basis back to $| \su \rangle, | \sd \rangle$ along spin-$z$ direction by the following unitary tranformation
\begin{align}
\bpm
\xi_{+} \\
\xi_{-}
\epm
=
\bpm
\cos (\theta/2) e^{-i\phi} &  \sin (\theta/2) \\
 -\sin (\theta/2) & \cos (\theta/2)e^{i\phi} 
\epm
\bpm
\xi_{\su} \\
\xi_{\sd}
\epm,
\end{align}
such that the transfer amplitude becomes 
\begin{align}
\mathcal{A}^M = \nu_L  \cdot \frac{ \arcsin(E_{Z, L}/\Delta_0)}{ 2\sqrt{\Delta^2_0 - E^2_{Z, L}} }  \times \Big [ ( \xi^2_{\sd} e^{i\phi} -  \xi^2_{\su} e^{-i\phi} ) \sin \theta + 2\xi_{\su} \xi_{\sd} \cos \theta \Big].
\end{align}

\section{Majorana supercurrent blockade}\label{app_sec:blockade}
The Hamiltonian of a time-reversal invariant superconductor with $s$-wave pairing symmetry can always be written in the following form
\begin{align}
H = H_0 + H_{sc} = \sum_{n} \Big \{  \varepsilon_n ( a\dg_n a_n + a\dg_{\nb} a_{\nb} ) + \Delta_0 ( a_{\nb} a_n + a\dg_n a\dg_{\nb}  ) \Big \},
\label{eq:ham_bcs}
\end{align}
where $a_n$ is the annihilation operator for a normal eigenstate of eigenenergy $\varepsilon_n$ and eigenfunction $\psi_n(x)$. $a_{\nb}$ is the annihilation operator for its time-reversed state which has an eigenenergy $\varepsilon_{\nb} = \varepsilon_n$ and eigenfunction $\psi_{\nb}(x)$. The relation between the original real-space electron operator $c(x)$ and the eigenstate operator $a_n$ is
\begin{align}
c_{\sigma}(x) = \sum_{n} \Big[ \psi_{n \sigma}(x) a_n + \psi_{\nb \sigma}(x) a_{\nb} \Big].
\label{eq:c}
\end{align}
On the other hand, since the Hamiltonian in Eq.~\eqref{eq:ham_bcs} is in the BCS form, we can expand the normal operators $a_n$ in terms of the Bogoliubov quasi-particle operators as
\begin{align}
& a_n = \tilde{u}_n \Gamma_{n} - \tilde{v}_n \Gamma\dg_{\nb}, \nn
& a_{\nb} = \tilde{u}_n \Gamma_{\nb} + \tilde{v}_n \Gamma\dg_{n},
\label{eq:a}
\end{align}
where $\Gamma\dg_{n}$ and $\Gamma\dg_{\nb}$ create Bogoliubov quasiparticles with excitation energy $E_{\nb} = E_n = \sqrt{\varepsilon^2_n + \Delta^2_0}$, and $\tilde{u}_n, \tilde{v}_n$ are BCS coherence factors with $\tilde{u}^2_n = 1/2 + \varepsilon_n / 2E_n = 1 - \tilde{v}^2_n$. Substituting Eq.~\eqref{eq:a} into Eq.~\eqref{eq:c}, we get 
\begin{align}
c_{\sigma}(x) = \sum_{n} \Big \{ \tilde{u}_n \big[  \psi_{n \sigma}(x) \Gamma_{n} + \psi_{\nb \sigma}(x) \Gamma_{\nb}  \big] + \tilde{v}_n \big[  \psi_{\nb \sigma}(x) \Gamma\dg_{n} -  \psi_{n \sigma}(x) \Gamma\dg_{\nb}  \big] \Big \}.
\label{eq:c_complete}
\end{align}
After plugging Eq.~\eqref{eq:c_complete} into Eqs.~\eqref{eq:A}, we have
\begin{align}
\mathcal{A}^M &=  \sum_{\eta, \sigma = \su \sd} \xi_{\eta}(x_R) \xi_{\sigma}(x_R) \sum_n \tilde{u}_n \tilde{v}_n \Big[ \frac{ \psi_{n \eta}(x_L)  \psi_{\nb \sigma}(x_L) }{E_n} - \frac{ \psi_{\nb \eta}(x_L)  \psi_{n \sigma}(x_L) }{E_{\nb}} \Big] \nn
&= \sum_{\eta, \sigma = \su \sd}  \xi_{\eta}(x_R) \xi_{\sigma}(x_R) \sum_n \frac{ \tilde{u}_n \tilde{v}_n }{E_n } \Big[  \psi_{n \eta}(x_L)  \psi_{\nb \sigma}(x_L) - \psi_{n \sigma}(x_L) \psi_{\nb \eta}(x_L)  \Big] \nn
&= \sum_{\eta, \sigma = \su \sd} \mathcal{A}^M_{\eta \sigma}.
\end{align}
For $\eta=\sigma$, 
\begin{align}
\mathcal{A}^M_{\sigma \sigma} \propto  [\psi_{n \sigma}(x_L)  \psi_{\nb \sigma}(x_L) - \psi_{n \sigma}(x_L) \psi_{\nb \sigma}(x_L) ] = 0.
\end{align}
For $\eta \neq \sigma$,
\begin{align}
\mathcal{A}^M_{\sd \su} + \mathcal{A}^M_{\su \sd} \propto  [\psi_{n \sd}(x_L)  \psi_{\nb \su}(x_L) - \psi_{n \su}(x_L) \psi_{\nb \sd}(x_L) ] + [\psi_{n \su}(x_L)  \psi_{\nb \sd}(x_L) - \psi_{n \sd}(x_L) \psi_{\nb \su}(x_L) ] = 0.
\end{align}
Thereby, Majorana-induced supercurrent is completely blockaded when the trivial superconducting lead is time-reversal invariant.

\end{document}